# Abnormally weak intervalley electron scattering in MoS$_2$ monolayer: insights from the matching between electron and phonon bands

Shiru Song,[†] Ji-Hui Yang[†‡]* and Xin-Gao Gong[†‡]

[†]Key Laboratory of Computational Physical Sciences (Ministry of Education), Institute of Computational Physics, Fudan University, Shanghai 200433, China

[‡]Shanghai Qi Zhi Institute, Shanghai 200230, China

**ABSTRACT:**

It is known that carrier mobility in layered semiconductors generally increases from two-dimension (2D) to three-dimension due to suppressed scattering channels resulting from decreased densities of electron and phonon states. In this work, we find an abnormal decrease of electron mobility from monolayer to bulk MoS$_2$. By carefully analyzing the scattering mechanisms, we can attribute such abnormality to the stronger intravalley scattering in the monolayer but weaker intervalley scattering caused by less intervalley scattering channels and weaker corresponding electron-phonon couplings compared to the bulk case. We show that, it is the matching between electronic band structure and phonon spectrum rather than their densities of electronic and phonon states that determines scattering channels. We propose, for the first time, the phonon-energy-resolved matching function to identify the intra- and inter-valley scattering channels. Furthermore, we show that multiple valleys do not necessarily lead to strong intervalley scattering if: (1) the scattering channels, which can be explicitly captured by the distribution of the matching function, are few due to the small matching between the corresponding electron and phonon bands; and/or (2) the multiple valleys are far apart in the reciprocal space and composed of out-of-plane orbitals so that the corresponding electron-phonon coupling strengths are weak. Consequently, the searching scope of high-mobility 2D materials can be reasonably enlarged using the matching function as useful guidance with the help of band edge orbital analysis.

# INTRODUCTION

Two-dimensional (2D) semiconductors without dangling bonds or short channel effects have shown great potential in the next generation of high-performance and low-power logic electronic devices, thus having attracted great interests in recent years.[1–7] However, one critical challenge hindering their applications is the relatively low carrier mobility, which has typical values of less than several hundred cm$^2$V$^{-1}$s$^{-1}$,[8–17] much smaller than that of silicon. Recently, several works have tried to reveal the origins of intrinsically low carrier mobility in 2D.[16,18] The generally accepted picture is that, with the layer number decreasing, both the densities of electronic states and low-energy phonon states increase, leading to the increased density of carrier intravalley scatterings and thus decreased mobility. However, experimental works reported that the mobility of MoS$_2$ first increases and then decreases with the increase of layer numbers.[19,20] They attributed the mobility increase to the change of Coulomb interaction distance[19] and surface roughness.[21] The mobility decrease was attributed to the Coulomb scattering due to charged impurities in thicker layers.[20] Apparently, the experimentally measured layer-dependence of carrier mobility are not fully consistent with current theories. Although partial reasons could be attributed to the environment-dependent factors such as defect and interfacial scattering, there might be intrinsic mechanisms that are uncovered, revealing which is not only of great importance for the applications but also will help designing reasonable guidance to high-mobility systems.

Note that, the physics related to the mobility can be complicated in general systems such as MoS$_2$, in which the electronic band structures might have multiple valleys for electrons or holes and phonon bands have non-linear optical modes. In this case, both intra- and inter-valley scattering channels could coexist and they can have different contributions to the scattering rates. While the intravalley scattering assisted by low-energy phonons can be well described by considering the densities of electron and phonon states, the intervalley scattering assisted by optical phonon modes is not well understood. Recent works proposed that nonpolar systems[22] with a steep and deep valley located at the Γ point to eliminate intervalley scattering[17] are strong candidates to have high mobility. However, do multiple valleys certainly lead to strong intervalley scattering and thus lead to low carrier mobility? If this is true, among known 2D semiconductors, very few systems can satisfy the above criteria. Otherwise, the searching scope of high-mobility 2D materials might be expanded. The critical thing is to identify intervalley scattering channels, understand the underlying mechanisms, and assess whether the intervalley scattering is likely to be strong or not before performing heavy electron-phonon coupling calculations.

In this work, using the first-principles calculation methods and considering all the electron-phonon couplings, we study the electron mobility in MoS$_2$ monolayer and bulk. We find that electron mobility decreases from the monolayer to the bulk, which is in contrast to the generally accepted picture. The abnormality is attributed to the stronger intravalley scattering but weaker intervalley scattering caused by less intervalley scattering channels and suppressed electron-phonon coupling in the monolayer compared to the bulk. We propose for the first time to use the phonon-energy-resolved matching function between electronic band structure and phonon spectrum rather than just the densities of electron and phonon states to identify potential channels. Furthermore, we find that multiple valleys do not necessarily lead to strong intervalley scattering, as we find that the intervalley scattering of the conduction band electron in the monolayer located at the K point is largely suppressed due to two reasons. One is the lack of channels due to the small match between electron and phonon bands, which is well captured in the distribution of the matching function. The other is that the corresponding electron-phonon coupling strengths are weak because the multiple valleys are far apart in the reciprocal space and composed of out-of-plane orbitals. Finally, we demonstrate that, the matching function together with the help of band edge orbital analysis can be used to judge whether a system has potentially high carrier mobility or not, thus speeding up the search of high-mobility 2D semiconductors in a broader scope.

**CALCULATION METHODS**

We use the ab initio Boltzmann transport equation to calculate carrier mobility in the self-energy relaxation time approximation (SERTA).[23] In this framework, the mobility is given by

$$\mu_{\alpha\beta} = \frac{-1}{V_{uc}n_c}\sum_n \int \frac{d^3k}{\Omega_{BZ}} \frac{\partial f^0_{n\mathbf{k}}}{\partial \varepsilon_{n\mathbf{k}}} v_{n\mathbf{k},\alpha} v_{n\mathbf{k},\beta} \tau_{n\mathbf{k}} \quad (1),$$

where $\alpha$, $\beta$ run over the three Cartesian directions, $n_c$ is the carrier concentration, $V_{uc}$ is the volume of unit cell, and $\Omega_{BZ}$ denotes the volume of the first Brillouin zone. The summation is over the band index n, and the integral is over the electron wavevectors **k** in the first Brillouin zone. $v_{n\mathbf{k},\alpha}$ is the band velocity for the Kohn-Sham state $|n\mathbf{k}\rangle$, and $f^0_{n\mathbf{k}}$ is the Fermi-Dirac equilibrium occupation at a given temperature. $\tau_{n\mathbf{k}}$ is the carrier scattering lifetime and its reciprocal value $\tau^{-1}_{n\mathbf{k}}$ is the scattering rate given by:

$$\tau_{n\mathbf{k}}^{-1} = \frac{2\pi}{\hbar}\sum_{m,v}\int\frac{d^3q}{\Omega_{BZ}}|g_{mnv}(\mathbf{k},\mathbf{q})|^2\big[(1-f^0_{m\mathbf{k+q}}+n_{\mathbf{q}v})\delta(\varepsilon_{n\mathbf{k}}-\varepsilon_{m\mathbf{k+q}}-\hbar\omega_{\mathbf{q}v}) +$$
$$(f^0_{m\mathbf{k+q}}+n_{\mathbf{q}v})\delta(\varepsilon_{n\mathbf{k}}-\varepsilon_{m\mathbf{k+q}}+\hbar\omega_{\mathbf{q}v})\big] \quad (2),$$

where $\mathbf{q}$ is a phonon wavevector, and the summation runs over the electron band index $m$ and phonon branch index $v$. $n_{\mathbf{q}v}$ is the Bose-Einstein distribution and $\omega_{\mathbf{q}v}$ is phonon of frequency. The electron-phonon matrix elements $g_{mnv}(\mathbf{k},\mathbf{q})$ are the amplitude for scattering from an initial state $|n\mathbf{k}\rangle$ to a final state $|m\mathbf{k}+\mathbf{q}\rangle$ via the emission or absorption of a phonon with indices $\mathbf{q}v$. The Dirac delta functions reflect the conservation of energy during the scattering process. Calculations of electron-phonon couplings and carrier mobilities are performed using the EPW code[24,25] based on first-principles calculations using the Quantum Espresso Package.[26,27] Detailed methods are provided in the Supporting Information.

## RESULTS AND DISCUSSIONS

As a typical system of transition metal dichalcogenides (TMD), MoS$_2$ has attracted great research interests.[28–34] Its electron mobility has been widely studied both experimentally and theoretically.[13,22,31,35–37] Our calculated electron mobility using SERTA is listed in the Table 1. As can be seen, our calculated electron mobility for monolayer MoS$_2$ is 158.3 cm$^2$V$^{-1}$s$^{-1}$, in good agreement with previous calculations.[18,22] Note that, from monolayer to bulk, the electron mobility decreases from 158.3 cm$^2$V$^{-1}$s$^{-1}$ to 97.1 cm$^2$V$^{-1}$s$^{-1}$. Compared to the models based on the electron and phonon density of states,[16,18] the behavior of electron mobility is kind of weird.

To understand the abnormal behavior of electron mobility, we consider two main mobility-determining factors in the SERTA—scattering rates and band velocities (and thus carrier effective masses). Note that, in Eq. (1), the derivative of the Fermi-Dirac distribution to the band energy is exponentially decayed away from the band edges in the case of relatively small carrier concentrations. Therefore, the scattering rates are mainly contributed from the band edge states. So we mainly focus on these states during the following discussions. As seen in Figs. 1a and 1b, the conduction band minimum (CBM) states in the monolayer and bulk are located at the K point and the Q point along $\Gamma - K$ line, respectively, both of which have multiple valleys (see Figs. 1c and 1d). Partial charge densities in Figs. 1e and 1f show that the CBM in the monolayer is composed of Mo $d_{z^2}$ orbitals while the CBM in the bulk is composed of Mo $d_{xy}, d_{x^2-y^2}, d_{z^2}$ orbitals.

As is known that, when multiple valleys present, intervalley scattering happens with the assist of optical phonon modes which has larger phonon energies. Therefore, to distinguish inter-and intra-valley scattering, we can use the phonon-spectra-decomposed scattering rate $\partial \tau_{n\mathbf{k}}^{-1}/\partial \omega$[38] defined according to:

$$\tau_{n\mathbf{k}}^{-1} = \int d\omega \, \partial \tau_{n\mathbf{k}}^{-1}/\partial \omega \quad (3),$$

where

$$\partial \tau_{n\mathbf{k}}^{-1}/\partial \omega = \frac{2\pi}{\hbar} \sum_{m,\nu} \int \frac{d^3 q}{\Omega_{BZ}} |g_{mn\nu}(\mathbf{k},\mathbf{q})|^2 \big[\big(1 - f^0_{m\mathbf{k}+\mathbf{q}} + n_{\mathbf{q}\nu}\big)\delta\big(\varepsilon_{n\mathbf{k}} - \varepsilon_{m\mathbf{k}+\mathbf{q}} - \hbar\omega_{\mathbf{q}\nu}\big) + \big(f^0_{m\mathbf{k}+\mathbf{q}} + n_{\mathbf{q}\nu}\big)\delta\big(\varepsilon_{n\mathbf{k}} - \varepsilon_{m\mathbf{k}+\mathbf{q}} + \hbar\omega_{\mathbf{q}\nu}\big)\big]\delta(\omega - \omega_{\mathbf{q}\nu}) \quad (4),$$

which reflects the density of scattering rate of an electron at an initial state $|n\mathbf{k}\rangle$ via the emission or absorption of a phonon with an energy of $\hbar\omega$. Then we further decompose $\partial \tau_{n\mathbf{k}}^{-1}/\partial \omega$ into two terms as the followings:

$$\partial \tau_{n\mathbf{k}}^{-1}/\partial \omega = 2\pi |g^*_{n\mathbf{k}}(\omega)|^2 F(\omega) \quad (5),$$

where,

$$F(\omega) = \frac{1}{\hbar} \sum_{m,\nu} \int \frac{d^3 q}{\Omega_{BZ}} \big[\big(1 - f^0_{m\mathbf{k}+\mathbf{q}} + n_{\mathbf{q}\nu}\big)\delta\big(\varepsilon_{n\mathbf{k}} - \varepsilon_{m\mathbf{k}+\mathbf{q}} - \hbar\omega_{\mathbf{q}\nu}\big) + \big(f^0_{m\mathbf{k}+\mathbf{q}} + n_{\mathbf{q}\nu}\big)\delta\big(\varepsilon_{n\mathbf{k}} - \varepsilon_{m\mathbf{k}+\mathbf{q}} + \hbar\omega_{\mathbf{q}\nu}\big)\big]\delta(\omega - \omega_{\mathbf{q}\nu}) \quad (6)$$

and

$$|g^*_{n\mathbf{k}}(\omega)|^2 = \frac{1}{2\pi} \left(\frac{\partial \tau_{n\mathbf{k}}^{-1}}{\partial \omega}\right)/F(\omega) \quad (7).$$

Note that, $F(\omega)$ relates the electronic structure to the phonon spectrum with several delta functions to evaluate the energy match between electron and phonon bands. Therefore, we define $F(\omega)$ as phonon-energy resolved matching function between electronic band structure and phonon spectrum and it characterizes the number of scattering channels. Once $F(\omega)$ is defined, phonon-energy resolved average EPC strength $g^*_{n\mathbf{k}}(\omega)$ can be obtained through Eq. (7).

Now we show how $F(\omega)$ and $g^*_{n\mathbf{k}}(\omega)$ can be used to understand the abnormal changes of electron mobility from monolayer to bulk MoS$_2$. As seen in Figs. 2(a) and 2(b), $\partial \tau_{n\mathbf{k}}^{-1}/\partial \omega$ in the monolayer has a significantly different distribution from that in the bulk, especially at the phonon-energy region of less than 250 cm$^{-1}$. On one hand, at the low-energy region (<50 cm$^{-1}$), $\partial \tau_{n\mathbf{k}}^{-1}/\partial \omega$ in the monolayer is several times larger than that in the bulk. Phonon-branch resolved scattering rate in Figs. 3a and 3b also shows that, the scattering rate due to low-energy phonons near the Γ point (intravalley scattering) is larger in the monolayer than that in the bulk. This is mainly because of larger $F(\omega)$ and thus more scattering channels in the monolayer due to the larger density of electronic states and low-energy phonon states.[18] On the other hand, in the energy region of 100–250 cm$^{-1}$, $\partial\tau_{n\mathbf{k}}^{-1}/\partial\omega$ in the bulk is several times larger. The phonon-branch resolved scattering rate in Figs. 2d indicates that the scattering is attributed to the intervalley scattering (because the contributions are mainly from phonons with momentum of $q\gg 0$ as seen in the Fig. 3b). This is because the CBM state in the bulk is located at the Q point along $\Gamma - K$ line and therefore has multiple valleys (see also Fig. 1d). The distribution of $F(\omega)$ shows that the matching function in this region is larger and broader in the bulk (see Figs. 2c and 2d), suggesting more scattering channels compared to the monolayer case. Together with the significant larger $g_{n\mathbf{k}}^*(\omega)$ (Figs. 2e and 2f), $\partial\tau_{n\mathbf{k}}^{-1}/\partial\omega$ has larger values in the region of 100–250 cm$^{-1}$ in the bulk. Besides, in the region of around 375 cm$^{-1}$ mainly corresponding to the intravalley scattering assisted by optical phonons, the distribution of $\partial\tau_{n\mathbf{k}}^{-1}/\partial\omega$ is also larger in the monolayer than the bulk because of more electronic and phonon density of states in the monolayer. Note that, it happens just by chance that the low-phonon-energy intravalley scattering in the monolayer is nearly equal to the high-phonon-energy intervalley scattering in the bulk. Considering that the electronic band velocities are larger in the monolayer with a smaller effective mass of 0.45 $m_e$ compared to 0.62 $m_e$ in the bulk, the monolayer has a larger mobility, which is in contrast to the simplified models considering just the densities of electron and phonon states.

Note that, the CBM states in the monolayer also have multiple valleys. However, the intervalley scattering is not significant compared to the bulk case. While we know that it's the different intervalley scattering effects that mainly contribute to the abnormal changes, we can also conclude that multiple valleys do NOT certainly lead to strong intervalley scattering. The different behaviors of intervalley scatterings can be understood from two aspects. On one hand, we see that, the matching functions in the monolayer are neither significantly large nor broadly distributed in the high-phonon-energy region around 100–250 cm$^{-1}$ compared to the bulk case, indicating that the intervalley scattering channels are less. On the other hand, the corresponding electron-phonon coupling strengths in the monolayer are also much weaker. This is because the CBM states in the monolayer are far apart in the reciprocal space and composed of Mo $d_{z^2}$ orbitals, which are out-of-plane distributed (see Fig. 1e). Instead, the CBM states in the bulk are mainly composed of Mo $d_{xy}$ and $d_{x^2-y^2}$ orbitals, which are in-plane distributed. As a result, the orbital couplings are weaker in the monolayer than in the bulk. Due to the above two points, the intervalley scatterings in monolayer $MoS_2$ are not strong and contribute little to the final scattering rates.

The weak intervalley scatterings between two equivalent K points in the honeycomb lattice are universal if the K valleys are constituent of out-of-plane orbitals as similar situations are found for the electron states in other TMDs as well as for the hole state in BN monolayer[39] (see the Supporting Information). However, if the multiple valleys are composed of the in-plane orbitals, such as the electron states in $MoS_2$ bulk and the hole state in $MoS_2$ monolayer (see the Supporting Information), the intervalley scattering will be strong, leading to relatively low mobility.

At last, we show that matching functions can also be used to qualitatively evaluate mobility, i.e., if we compare the matching function of a system to that of $MoS_2$, we can know whether the system has larger or smaller mobility than $MoS_2$. To demonstrate this, we compare the matching functions for the CBM states of $MoS_2$, $WS_2$, $MoSe_2$ and black phosphorus as well as the VBM state of BN (see the Supporting Information. Note that, all of these states have weak intervalley scatterings). We plot the calculated carrier mobility (calculated by different groups) versus the maxima of the matching functions (calculated in this work) by considering the effects of carrier effective masses. As shown in Fig. 4, we can see that the mobility indeed has strong positive correlations with the reciprocals of the maxima of the matching functions. Therefore, we can use the matching functions together with considerations of band edge orbital analysis to find potentially high carrier mobility in systems with multiple valleys. In fact, we indeed identify some systems with exceptionally high mobility which be reported elsewhere.[40]

**CONCLUSION**

In summary, we have proposed to use the phonon-energy resolved matching functions between electronic band structure and phonon spectrum rather than just densities of electron and phonon states to identify the potential scattering channels. With the help of matching functions, we have explained the abnormal decrease of electron mobility from monolayer to bulk $MoS_2$ due to less intervalley scattering channels and suppressed corresponding electron-phonon couplings. We have further shown that multiple valleys do not necessarily lead to strong intervalley scattering if the matching functions are small and/or the valley orbitals are far apart in the reciprocal space and out-of-plane distributed. Finally, we have demonstrated the matching functions together with band edge analyses can be used to qualitatively assess carrier mobility and help identifying potentially high-mobility systems.

## ASSOCIATED CONTENT

**Supporting Information**

Supporting information for this article are available at XXXX.

Detailed calculation methods and band edge valleys and partial charge of the hole in $MoS_2$ monolayer, and other supporting data.

Fig. S1. Band edge valleys and partial charge of the hole in $MoS_2$ monolayer.

Fig. S2. The phonon-energy resolved hole scattering rates, matching functions, average electron-phonon coupling strength, and phonon-branch resolved scattering in monolayer $MoS_2$.

Fig. S3. The phonon-energy resolved electron scattering rates, matching functions, average electron-phonon coupling strength, and phonon-branch resolved scattering in monolayer $MoSe_2$.

Fig. S4. The phonon-energy resolved electron scattering rates, matching functions, average electron-phonon coupling strength, and phonon-branch resolved scattering in monolayer $WS_2$.

Fig. S5. The phonon-energy resolved electron scattering rates, matching functions, average electron-phonon coupling strength, and phonon-branch resolved scattering in the monolayer black phosphorus.

Fig. S6. The phonon-energy resolved hole scattering rates, matching functions, average electron-phonon coupling strength, and phonon-branch resolved scattering in the monolayer BN.

## AUTHOR INFORMATION

**Corresponding Author**
Email: jhyang04@fudan.edu.cn**Author contributions**

J.-H. Y. conceived the idea, S.-R. S., J.-H. Y. and X.-G. G. analyzed the data, and wrote the paper, S.-R. S. performed the calculations.

**Notes**

The authors declare that they have no competing interests.

**ACKNOWLEGEMENTS**

This work was supported in part by the National Natural Science Foundation of China (Grant No. 12188101, 61904035, 11974078, 11991061) and the Shanghai Sailing Program(19YF1403100). The supercomputing time was sponsored by the Fudan University High-End Computing Center.

**Table**

**Table 1.** Calculated effective masses and mobility for electrons in $MoS_2$ monolayer and bulk. ∥ denotes the in-plane case, and ⊥ denotes the out-plane case. $m_e$ is the electron rest mass.

|  | Mass($m_e$) | Mobility(cm$^2$/Vs) |
| --- | --- | --- |
| monolayer | 0.50 | 158.3 |
| bulk | 0.62(∥),0.61(⊥) | 97.1(∥),111.5(⊥) |

**FIGURES**

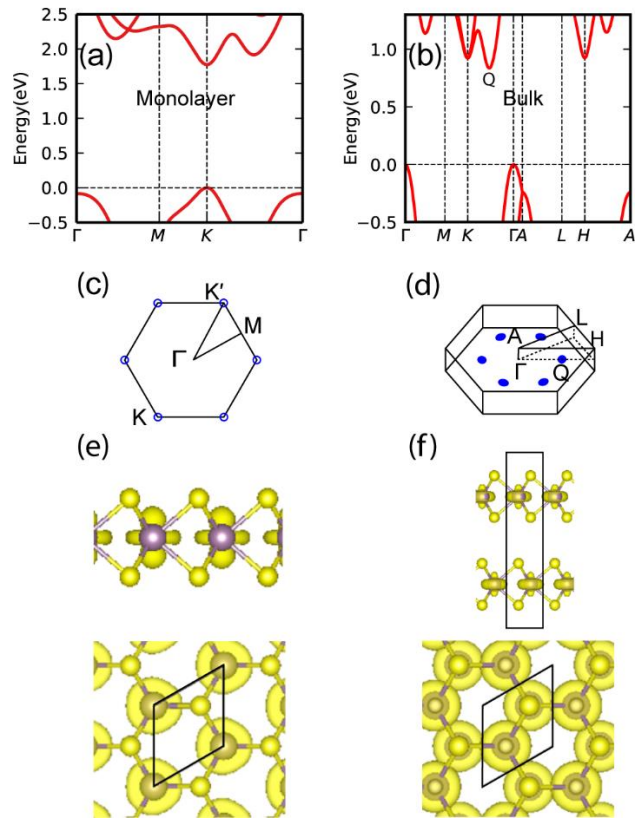

**Figure 1.** The band structures and CBM valleys of MoS$_2$ monolayer and bulk. (a) and (b) Band structure of monolayer and bulk MoS$_2$. (c) The CBM valleys in monolayer MoS$_2$. (d) The CBM valleys in bulk MoS$_2$. (e) Top and side view of the CBM partial charge densities in monolayer MoS$_2$. (f) Top and side view of the CBM partial charge densities in bulk MoS$_2$.

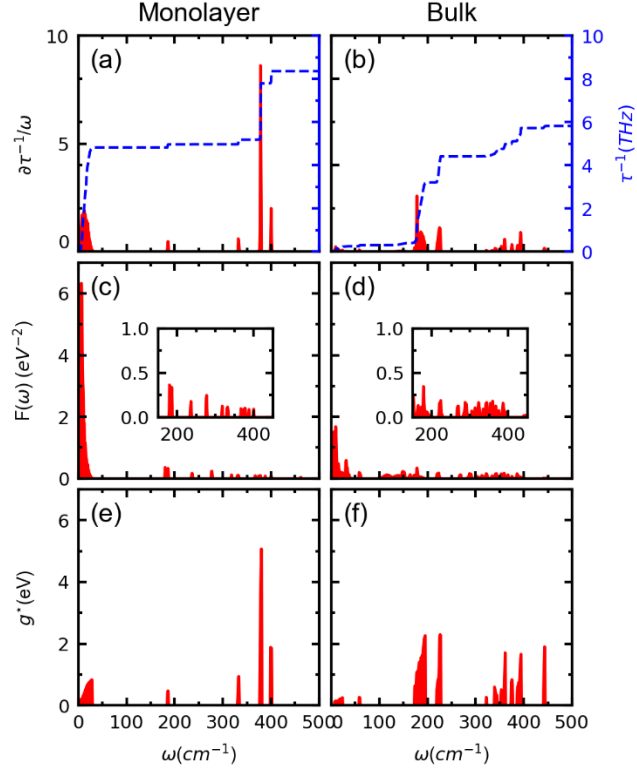

**Figure 2.** The phonon-energy resolved electron scattering rates, matching functions, and average electron-phonon coupling strength in monolayer and bulk $MoS_2$. (a) and (b) The phonon-spectra-decomposed electron scattering rate. (c) and (d) The phonon-energy resolved matching functions between electronic band structure and phonon spectrum. (e) and (f) The phonon-energy resolved average electron-phonon coupling strength for monolayer and bulk $MoS_2$. Note that, the blue lines in (a) and (b) indicate the integrated scattering rates.

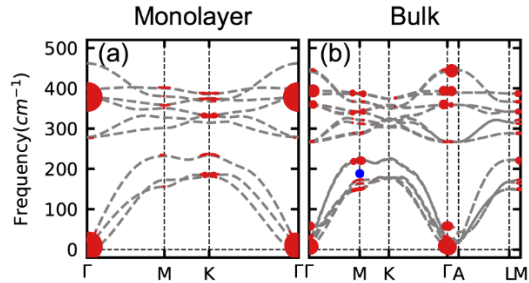

**Figure 3.** The phonon-branch resolved scattering rates in monolayer and bulk $MoS_2$. (a) and (b) The phonon-branch resolved scattering rates for electrons in $MoS_2$ monolayer and bulk, respectively. Note that, the sizes of the dots are proportional to the scattering rates assisted by the corresponding phonon modes. For the electron case in bulk $MoS_2$, the large contribution to the scattering rates around 200 cm$^{-1}$ is attributed to the phonon modes deviating from the high symmetry lines in the Brillouin zone, which have been denoted by the blue dot.

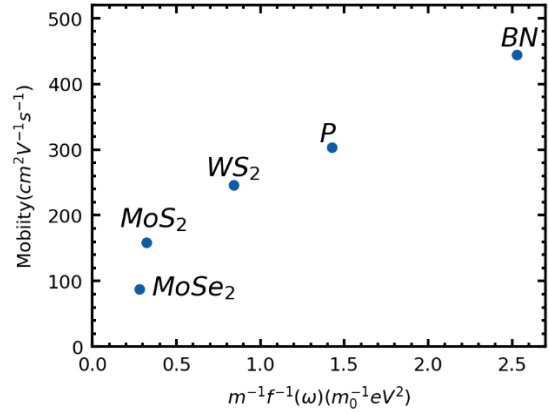

**Figure 4.** Carrier mobility versus the reciprocals of the maxima of the matching functions timing the carrier effective masses. Note that, the mobilities for CBM states of $MoS_2$, $WS_2$, $MoSe_2$[18,22] and black phosphorus[18] as well as the VBM state of BN[39] are all taken from previous works from different groups while the $F(\omega)$'s are calculated in this work. The strong positive correlations between the carrier mobility and the reciprocals of the maxima of the matching functions can be clearly seen.

**References:**

(1) Akinwande, D.; Petrone, N.; Hone, J. Two-Dimensional Flexible Nanoelectronics. *Nat. Commun.* **2014**, *5* (1), 5678. https://doi.org/10/f6ttr5.
(2) Novoselov, K. S.; Jiang, D.; Schedin, F.; Booth, T. J.; Khotkevich, V. V.; Morozov, S. V.; Geim, A. K. Two-Dimensional Atomic Crystals. *Proc. Natl. Acad. Sci. U. S. A.* **2005**, *102* (30), 10451–10453. https://doi.org/10.1073/pnas.0502848102.
(3) Akinwande, D.; Huyghebaert, C.; Wang, C.-H.; Serna, M. I.; Goossens, S.; Li, L.-J.; Wong, H.-S. P.; Koppens, F. H. L. Graphene and Two-Dimensional Materials for Silicon Technology. *Nature* **2019**, *573* (7775), 507–518. https://doi.org/10/gf8v3z.
(4) Li, M.-Y.; Su, S.-K.; Wong, H.-S. P.; Li, L.-J. How 2D Semiconductors Could Extend Moore's Law. *Nature* **2019**, *567* (7747), 169–170. https://doi.org/10/gfwsz7.
(5) Lin, Z.; Liu, Y.; Halim, U.; Ding, M.; Liu, Y.; Wang, Y.; Jia, C.; Chen, P.; Duan, X.; Wang, C.; Song, F.; Li, M.; Wan, C.; Huang, Y.; Duan, X. Solution-Processable 2D Semiconductors for High-Performance Large-Area Electronics. *Nature* **2018**, *562* (7726), 254–258. https://doi.org/10/gd878s.
(6) Polyushkin, D. K.; Wachter, S.; Mennel, L.; Paur, M.; Paliy, M.; Iannaccone, G.; Fiori, G.; Neumaier, D.; Canto, B.; Mueller, T. Analogue Two-Dimensional Semiconductor Electronics. *Nat. Electron.* **2020**, *3* (8), 486–491. https://doi.org/10/gpfqr6.
(7) Cao, Y.; Fatemi, V.; Fang, S.; Watanabe, K.; Taniguchi, T.; Kaxiras, E.; Jarillo-Herrero, P. Unconventional Superconductivity in Magic-Angle Graphene Superlattices. *Nature* **2018**. https://doi.org/10.1038/nature26160.
(8) Radisavljevic, B.; Radenovic, A.; Brivio, J.; Giacometti, V.; Kis, A. Single-Layer $MoS_2$ Transistors. *Nat. Nanotechnol.* **2011**, *6* (3), 147–150. https://doi.org/10.1038/nnano.2010.279.
(9) Chhowalla, M.; Liu, Z.; Zhang, H. Two-Dimensional Transition Metal Dichalcogenide (TMD) Nanosheets. *Chem. Soc. Rev.* **2015**, *44* (9), 2584–2586. https://doi.org/10.1039/C5CS90037A.
(10) Liu, Y.; Duan, X.; Huang, Y.; Duan, X. Two-Dimensional Transistors beyond Graphene and TMDCs. *Chem. Soc. Rev.* **2018**, *47* (16), 6388–6409. https://doi.org/10/gd234g.
(11) Li, S.-L.; Tsukagoshi, K.; Orgiu, E.; Samorì, P. Charge Transport and Mobility Engineering in Two-Dimensional Transition Metal Chalcogenide Semiconductors. *Chem. Soc. Rev.* **2015**, *45* (1), 118–151. https://doi.org/10/ghb3gn.
(12) Li, L.; Yu, Y.; Ye, G. J.; Ge, Q.; Ou, X.; Wu, H.; Feng, D.; Chen, X. H.; Zhang, Y. Black Phosphorus Field-Effect Transistors. *Nat. Nanotechnol.* **2014**, *9* (5), 372–377. https://doi.org/10/gd24b7.
(13) Kaasbjerg, K.; Thygesen, K. S.; Jacobsen, K. W. Phonon-Limited Mobility in n-Type Single-Layer $MoS_2$ from First Principles. *Phys. Rev. B* **2012**, *85* (11), 115317. https://doi.org/10/ghhpr8.
(14) Gunst, T.; Markussen, T.; Stokbro, K.; Brandbyge, M. First-Principles Method for Electron-Phonon Coupling and Electron Mobility: Applications to Two-Dimensional Materials. *Phys. Rev. B* **2016**, *93* (3), 035414. https://doi.org/10/gf668r.
(15) Sohier, T.; Campi, D.; Marzari, N.; Gibertini, M. Mobility of Two-Dimensional Materials from First Principles in an Accurate and Automated Framework. *Phys. Rev. Mater.* **2018**, *2* (11), 114010. https://doi.org/10/ghjrpz.
(16) Li, W.; Poncé, S.; Giustino, F. Dimensional Crossover in the Carrier Mobility of Two-Dimensional Semiconductors: The Case of InSe. *Nano Lett.* **2019**, *19* (3), 1774–1781. https://doi.org/10/gf2wsf.
(17) Cheng, L.; Zhang, C.; Liu, Y. The Optimal Electronic Structure for High-Mobility 2D Semiconductors: Exceptionally High Hole Mobility in 2D Antimony. *J. Am. Chem. Soc.* **2019**, *141* (41), 16296–16302. https://doi.org/10/gncws7.
(18) Cheng, L.; Zhang, C.; Liu, Y. Why Two-Dimensional Semiconductors Generally Have Low Electron Mobility. *Phys. Rev. Lett.* **2020**, *125* (17), 177701. https://doi.org/10.1103/PhysRevLett.125.177701.
(19) Li, S.-L.; Wakabayashi, K.; Xu, Y.; Nakaharai, S.; Komatsu, K.; Li, W.-W.; Lin, Y.-F.; Aparecido-Ferreira, A.; Tsukagoshi, K. Thickness-Dependent Interfacial Coulomb Scattering in Atomically Thin Field-Effect Transistors. *Nano Lett.* **2013**, *13* (8), 3546–3552. https://doi.org/10.1021/nl4010783.
(20) Lin, M.-W.; Kravchenko, I. I.; Fowlkes, J.; Li, X.; Puretzky, A. A.; Rouleau, C. M.; Geohegan, D. B.; Xiao, K. Thickness-Dependent Charge Transport in Few-Layer $MoS_2$ Field-Effect Transistors. *Nanotechnology* **2016**, *27* (16), 165203. https://doi.org/10/gm2ztm.
(21) Kovalenko, K. L.; Kozlovskiy, S. I.; Sharan, N. N. Electron Mobility in Molybdenum Disulfide: From Bulk to Monolayer. *Phys. Status Solidi B* **2020**, *257* (5), 1900635. https://doi.org/10/ghjrn5.
(22) Cheng, L.; Liu, Y. What Limits the Intrinsic Mobility of Electrons and Holes in Two Dimensional Metal Dichalcogenides? *J. Am. Chem. Soc.* **2018**, *140* (51), 17895–17900. https://doi.org/10/gg73sm.

(23) Poncé, S.; Li, W.; Reichardt, S.; Giustino, F. First-Principles Calculations of Charge Carrier Mobility and Conductivity in Bulk Semiconductors and Two-Dimensional Materials. *Rep. Prog. Phys.* **2020**, *83* (3), 036501. https://doi.org/10/ggr8v7.

(24) Poncé, S.; Margine, E. R.; Verdi, C.; Giustino, F. EPW: Electron–Phonon Coupling, Transport and Superconducting Properties Using Maximally Localized Wannier Functions. *Comput. Phys. Commun.* **2016**, *209*, 116–133. https://doi.org/10/ggfx46.

(25) Giustino, F.; Cohen, M. L.; Louie, S. G. Electron-Phonon Interaction Using Wannier Functions. *Phys. Rev. B* **2007**, *76* (16), 165108. https://doi.org/10.1103/PhysRevB.76.165108.

(26) Giannozzi, P.; Baroni, S.; Bonini, N.; Calandra, M.; Car, R.; Cavazzoni, C.; Davide Ceresoli; Chiarotti, G. L.; Cococcioni, M.; Dabo, I.; Corso, A. D.; Gironcoli, S. de; Fabris, S.; Fratesi, G.; Gebauer, R.; Gerstmann, U.; Gougoussis, C.; Anton Kokalj; Lazzeri, M.; Martin-Samos, L.; Marzari, N.; Mauri, F.; Mazzarello, R.; Stefano Paolini; Pasquarello, A.; Paulatto, L.; Sbraccia, C.; Scandolo, S.; Sclauzero, G.; Seitsonen, A. P.; Smogunov, A.; Umari, P.; Wentzcovitch, R. M. QUANTUM ESPRESSO: A Modular and Open-Source Software Project for Quantum Simulations of Materials. *J. Phys. Condens. Matter* **2009**, *21* (39), 395502. https://doi.org/10.1088/0953-8984/21/39/395502.

(27) Giannozzi, P.; Andreussi, O.; Brumme, T.; Bunau, O.; Nardelli, M. B.; Calandra, M.; Car, R.; Cavazzoni, C.; Ceresoli, D.; Cococcioni, M.; Colonna, N.; Carnimeo, I.; Corso, A. D.; Gironcoli, S. de; Delugas, P.; DiStasio, R. A.; Ferretti, A.; Floris, A.; Fratesi, G.; Fugallo, G.; Gebauer, R.; Gerstmann, U.; Giustino, F.; Gorni, T.; Jia, J.; Kawamura, M.; Ko, H.-Y.; Kokalj, A.; Küçükbenli, E.; Lazzeri, M.; Marsili, M.; Marzari, N.; Mauri, F.; Nguyen, N. L.; Nguyen, H.-V.; Otero-de-la-Roza, A.; Paulatto, L.; Poncé, S.; Rocca, D.; Sabatini, R.; Santra, B.; Schlipf, M.; Seitsonen, A. P.; Smogunov, A.; Timrov, I.; Thonhauser, T.; Umari, P.; Vast, N.; Wu, X.; Baroni, S. Advanced Capabilities for Materials Modelling with Quantum ESPRESSO. *J. Phys. Condens. Matter* **2017**, *29* (46), 465901. https://doi.org/10/gf6627.

(28) Hung, T. Y. T.; Camsari, K. Y.; Zhang, S.; Upadhyaya, P.; Chen, Z. Direct Observation of Valley-Coupled Topological Current in MoS$_2$. *Sci. Adv. 5* (4), eaau6478. https://doi.org/10/gpf6q5.

(29) Roch, J. G.; Froehlicher, G.; Leisgang, N.; Makk, P.; Watanabe, K.; Taniguchi, T.; Warburton, R. J. Spin-Polarized Electrons in Monolayer MoS$_2$. *Nat. Nanotechnol.* **2019**, *14* (5), 432–436. https://doi.org/10/gfw2p2.

(30) Jiang, J.; Chen, Z.; Hu, Y.; Xiang, Y.; Zhang, L.; Wang, Y.; Wang, G.-C.; Shi, J. Flexo-Photovoltaic Effect in MoS$_2$. *Nat. Nanotechnol.* **2021**, *16* (8), 894–901. https://doi.org/10/gpb6mm.

(31) Sebastian, A.; Pendurthi, R.; Choudhury, T. H.; Redwing, J. M.; Das, S. Benchmarking Monolayer MoS$_2$ and WS$_2$ Field-Effect Transistors. *Nat. Commun.* **2021**, *12* (1), 693. https://doi.org/10/gjrp66.

(32) Mak, K. F.; Lee, C.; Hone, J.; Shan, J.; Heinz, T. F. Atomically Thin MoS$_2$: A New Direct-Gap Semiconductor. *Phys. Rev. Lett.* **2010**, *105* (13), 136805. https://doi.org/10.1103/PhysRevLett.105.136805.

(33) Splendiani, A.; Sun, L.; Zhang, Y.; Li, T.; Kim, J.; Chim, C.-Y.; Galli, G.; Wang, F. Emerging Photoluminescence in Monolayer MoS$_2$. *Nano Lett.* **2010**, *10* (4), 1271–1275. https://doi.org/10.1021/nl903868w.

(34) Bhimanapati, G. R.; Lin, Z.; Meunier, V.; Jung, Y.; Cha, J.; Das, S.; Xiao, D.; Son, Y.; Strano, M. S.; Cooper, V. R.; Liang, L.; Louie, S. G.; Ringe, E.; Zhou, W.; Kim, S. S.; Naik, R. R.; Sumpter, B. G.; Terrones, H.; Xia, F.; Wang, Y.; Zhu, J.; Akinwande, D.; Alem, N.; Schuller, J. A.; Schaak, R. E.; Terrones, M.; Robinson, J. A. Recent Advances in Two-Dimensional Materials beyond Graphene. *ACS Nano* **2015**, *9* (12), 11509–11539. https://doi.org/10/f75fsd.

(35) Radisavljevic, B.; Kis, A. Mobility Engineering and a Metal–Insulator Transition in Monolayer MoS$_2$. *Nat. Mater.* **2013**, *12* (9), 815–820. https://doi.org/10/f47jsx.

(36) Cai, Y.; Zhang, G.; Zhang, Y.-W. Polarity-Reversed Robust Carrier Mobility in Monolayer MoS$_2$ Nanoribbons. *J. Am. Chem. Soc.* **2014**, *136* (17), 6269–6275. https://doi.org/10/f5z5n4.

(37) Yu, Z.; Ong, Z.-Y.; Pan, Y.; Cui, Y.; Xin, R.; Shi, Y.; Wang, B.; Wu, Y.; Chen, T.; Zhang, Y.-W.; Zhang, G.; Wang, X. Realization of Room-Temperature Phonon-Limited Carrier Transport in Monolayer MoS$_2$ by Dielectric and Carrier Screening. *Adv. Mater.* **2016**, *28* (3), 547–552. https://doi.org/10/f3j5t4.

(38) Poncé, S.; Schlipf, M.; Giustino, F. Origin of Low Carrier Mobilities in Halide Perovskites. *ACS Energy Lett.* **2019**, *4* (2), 456–463. https://doi.org/10/gmd3f5.

(39) Khatami, M. M.; Van de Put, M. L.; Vandenberghe, W. G. First-Principles Study of Electronic Transport in Germanane and Hexagonal Boron Nitride. *Phys. Rev. B* **2021**, *104* (23), 235424. https://doi.org/10/gntjz2.

(40) Unpublished.

SUPPORTING INFORMATION for:

# Identifying intra- and inter-valley scattering channels from the match between electron bands and phonon spectrum: the case of MoS$_2$

Shiru Song,[†] Ji-Hui Yang[†‡]* and Xin-Gao Gong,[†‡]

[†]Key Laboratory of Computational Physical Sciences (Ministry of Education), Institute of Computational Physics, Fudan University, Shanghai 200433, China
[‡]Shanghai Qi Zhi Institute, Shanghai 200230, China
Email: jhyang04@fudan.edu.cn

Detailed calculation methods and band edge valleys and partial charge of the hole in MoS$_2$ monolayer, and other supporting data.

Fig. S1. Band edge valleys and partial charge of the hole in MoS$_2$ monolayer.

Fig. S2. The phonon-energy resolved hole scattering rates, matching functions, average electron-phonon coupling strength, and phonon-branch resolved scattering in monolayer MoS$_2$.

Fig. S3. The phonon-energy resolved electron scattering rates, matching functions, average electron-phonon coupling strength, and phonon-branch resolved scattering in monolayer MoSe$_2$.

Fig. S4. The phonon-energy resolved electron scattering rates, matching functions, average electron-phonon coupling strength, and phonon-branch resolved scattering in monolayer WS$_2$.

Fig. S5. The phonon-energy resolved electron scattering rates, matching functions, average electron-phonon coupling strength, and phonon-branch resolved scattering in the monolayer black phosphorus.

Fig. S6. The phonon-energy resolved hole scattering rates, matching functions, average electron-phonon coupling strength, and phonon-branch resolved scattering in the monolayer BN.

**Supplementary Note S1: Computation Methods**

We perform first-principles calculations using the Quantum Espresso Package[1,2] with the norm-conserving pseudopotentials[3] and the Perdew-Burke-Ernzerhof (PBE) exchange-correlation functional[4]. The van der Waals interaction in $MoS_2$ bulk is considered with the nonlocal functional rVV10[5]. The kinetic energy cutoffs for wave functions and charge density are set to 80 and 320 Ry, respectively. The atomic coordinates for bulk $MoS_2$ are optimized using a 36 × 36 × 8 k mesh until the force acting on each atom becomes less than 0.0001 Ry/Bohr. The monolayer $MoS_2$ is described using a vacuum-slab model. The length of the cell in the out-of-plane direction is 20 Å for monolayer $MoS_2$. The Brillouin zone is sampled using a 36 ×36 Monkhorst-Pack mesh. Calculations of electron-phonon couplings and carrier mobilities are performed using the EPW code[6,7]. For bulk $MoS_2$, the electron-phonon matrix elements are initially computed on a 18 × 18 × 4 electronic grid and a 9 × 9 × 2 phonon grid, using density functional perturbation theory[8] (DFPT). Then the electron-phonon matrix elements are subsequently interpolated onto fine grids, a 90 × 90 × 21 electronic grid and a 90 × 90 × 21 phonon grid, using maximally localized Wannier functions[9,10]. For monolayer, the electron-phonon matrix elements are initially computed on a 18 × 18 electronic grid and a 9 × 9 phonon grid, which are subsequently interpolated onto fine grids, i.e., 300 × 300 electronic grid and a 300 × 300 phonon grid. The carrier concentration for computing the mobilities is set to $10^{16}$ cm$^{-3}$, and the temperature is set to 300 K for all systems.

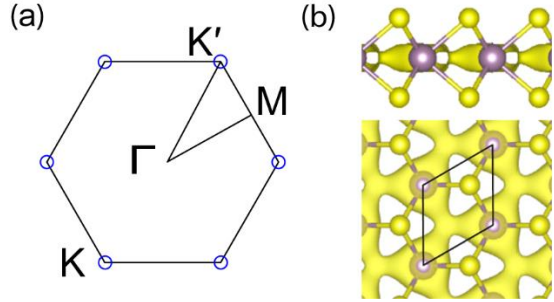

**Figure S1.** Band edge valleys and partial charge of the hole in MoS$_2$ monolayer. (a) The VBM valleys in monolayer MoS$_2$. (b) Top and side view of the VBM partial charge densities.

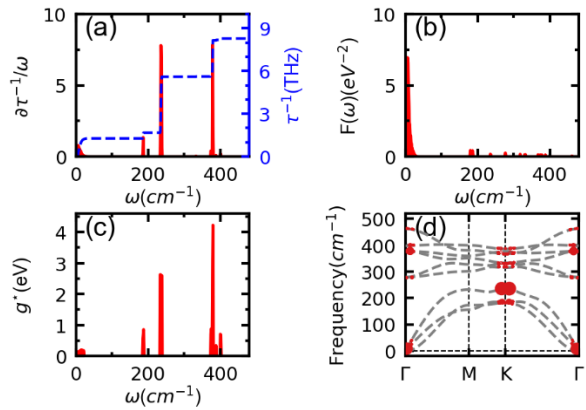

**Figure S2.** The phonon-energy resolved hole scattering rates, matching functions, average electron-phonon coupling strength, and phonon-branch resolved scattering in monolayer MoS$_2$. (a) The phonon-spectra-decomposed hole scattering rate. (b) The phonon-energy resolved matching functions between electronic band structure and phonon spectrum. (c) The phonon-energy resolved average electron-phonon coupling strength for monolayer MoS$_2$. (d) The phonon-branch resolved scattering rates for holes in the MoS$_2$ monolayer. Note that, the blue lines in (a) indicate the integrated scattering rates and the sizes of the dots in (d) are proportional to the scattering rates assisted by the corresponding phonon modes.

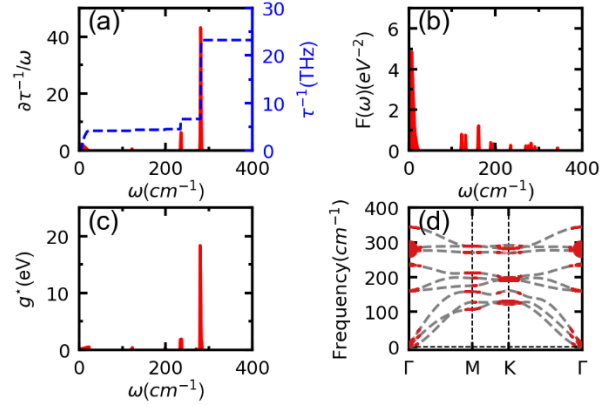

**Figure S3.** The phonon-energy resolved electron scattering rates, matching functions, average electron-phonon coupling strength, and phonon-branch resolved scattering in monolayer $MoSe_2$. (a) The phonon-spectra-decomposed electron scattering rate. (b) The phonon-energy resolved matching functions between electronic band structure and phonon spectrum. (c) The phonon-energy resolved average electron-phonon coupling strength for monolayer $MoSe_2$. (d) The phonon-branch resolved scattering rates for electrons in the $MoSe_2$ monolayer.

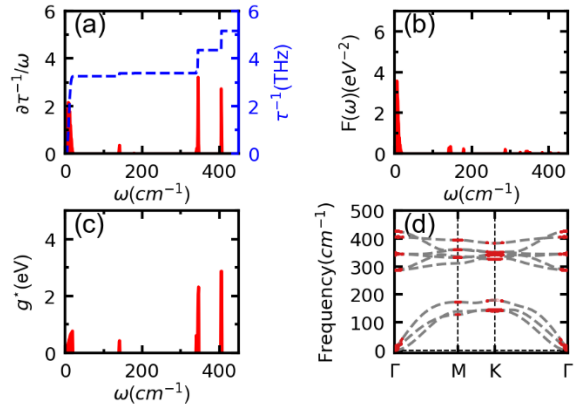

**Figure S4.** The phonon-energy resolved electron scattering rates, matching functions, average electron-phonon coupling strength, and phonon-branch resolved scattering in monolayer $WS_2$. (a) The phonon-spectra-decomposed electron scattering rate. (b) The phonon-energy resolved matching functions between electronic band structure and phonon spectrum. (c) The phonon-energy resolved average electron-phonon coupling strength for monolayer $WS_2$. (d) The phonon-branch resolved scattering rates for electrons in the $WS_2$ monolayer.

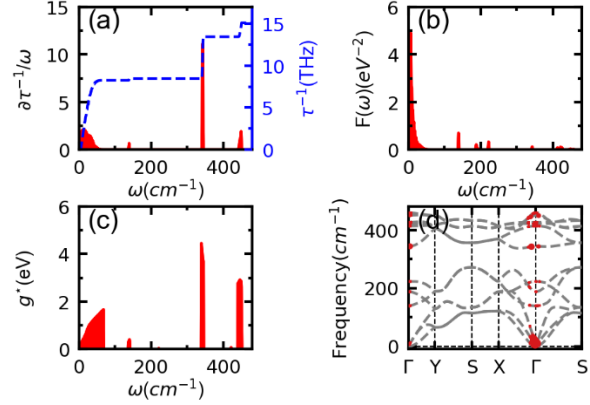

**Figure S5.** The phonon-energy resolved electron scattering rates, matching functions, average electron-phonon coupling strength, and phonon-branch resolved scattering in the monolayer black phosphorus. (a) The phonon-spectra-decomposed electron scattering rate. (b) The phonon-energy resolved matching functions between electronic band structure and phonon spectrum. (c) The phonon-energy resolved average electron-phonon coupling strength for monolayer black phosphorus. (d) The phonon-branch resolved scattering rates for electrons in the black phosphorus monolayer.

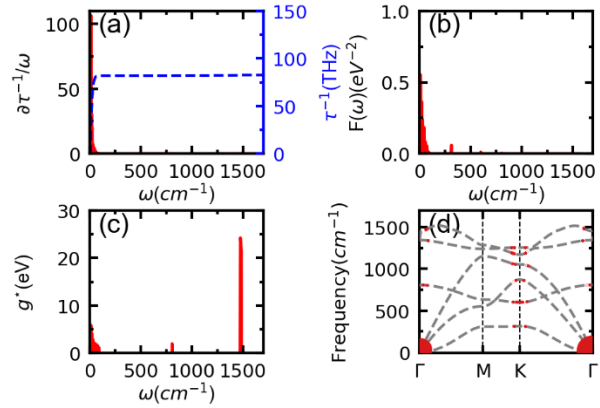

**Figure S6.** The phonon-energy resolved hole scattering rates, matching functions, average electron-phonon coupling strength, and phonon-branch resolved scattering in the monolayer BN. (a) The phonon-spectra-decomposed hole scattering rate. (b) The phonon-energy resolved matching functions between electronic band structure and phonon spectrum. (c) The phonon-energy resolved average electron-phonon coupling strength for monolayer BN. (d) The phonon-branch resolved scattering rates for holes in the BN monolayer.

**REFERENCES:**
(1) Giannozzi, P.; Baroni, S.; Bonini, N.; Calandra, M.; Car, R.; Cavazzoni, C.; Davide Ceresoli; Chiarotti, G. L.; Cococcioni, M.; Dabo, I.; Corso, A. D.; Gironcoli, S. de; Fabris, S.; Fratesi, G.; Gebauer, R.; Gerstmann, U.; Gougoussis, C.; Anton Kokalj; Lazzeri, M.; Martin-Samos, L.; Marzari, N.; Mauri, F.; Mazzarello, R.; Stefano Paolini; Pasquarello, A.; Paulatto, L.; Sbraccia, C.; Scandolo, S.; Sclauzero, G.; Seitsonen, A. P.; Smogunov, A.; Umari, P.; Wentzcovitch, R. M. QUANTUM ESPRESSO: A Modular and Open-Source Software Project for Quantum Simulations of Materials. *J. Phys.: Condens. Matter* **2009**, *21* (39), 395502. https://doi.org/10.1088/0953-8984/21/39/395502.
(2) Giannozzi, P.; Andreussi, O.; Brumme, T.; Bunau, O.; Nardelli, M. B.; Calandra, M.; Car, R.; Cavazzoni, C.; Ceresoli, D.; Cococcioni, M.; Colonna, N.; Carnimeo, I.; Corso, A. D.; Gironcoli, S. de; Delugas, P.; DiStasio, R. A.; Ferretti, A.; Floris, A.; Fratesi, G.; Fugallo, G.; Gebauer, R.; Gerstmann, U.; Giustino, F.; Gorni, T.; Jia, J.; Kawamura, M.; Ko, H.-Y.; Kokalj, A.; Küçükbenli, E.; Lazzeri, M.; Marsili, M.; Marzari, N.; Mauri, F.; Nguyen, N. L.; Nguyen, H.-V.; Otero-de-la-Roza, A.; Paulatto, L.; Poncé, S.; Rocca, D.; Sabatini, R.; Santra, B.; Schlipf, M.; Seitsonen, A. P.; Smogunov, A.; Timrov, I.; Thonhauser, T.; Umari, P.; Vast, N.; Wu, X.; Baroni, S. Advanced Capabilities for Materials Modelling with Quantum ESPRESSO. *J. Phys.: Condens. Matter* **2017**, *29* (46), 465901. https://doi.org/10/gf6627.
(3) Hamann, D. R. Optimized Norm-Conserving Vanderbilt Pseudopotentials. *Phys. Rev. B* **2013**, *88* (8), 085117. https://doi.org/10.1103/PhysRevB.88.085117.
(4) Perdew, J. P.; Burke, K.; Ernzerhof, M. Generalized Gradient Approximation Made Simple. *Phys. Rev. Lett.* **1996**, *77* (18), 3865–3868. https://doi.org/10.1103/PhysRevLett.77.3865.
(5) Sabatini, R.; Gorni, T.; de Gironcoli, S. Nonlocal van Der Waals Density Functional Made Simple and Efficient. *Phys. Rev. B* **2013**, *87* (4), 041108. https://doi.org/10/gftwdt.
(6) Poncé, S.; Margine, E. R.; Verdi, C.; Giustino, F. EPW: Electron-Phonon Coupling, Transport and Superconducting Properties Using Maximally Localized Wannier Functions. *Comput. Phys. Commun.* **2016**, *209*, 116–133. https://doi.org/10/ggfx46.
(7) Giustino, F.; Cohen, M. L.; Louie, S. G. Electron-Phonon Interaction Using Wannier Functions. *Phys. Rev. B* **2007**, *76* (16), 165108. https://doi.org/10.1103/PhysRevB.76.165108.
(8) Giustino, F. Electron-Phonon Interactions from First Principles. *Rev. Mod. Phys.* **2017**, *89* (1), 015003. https://doi.org/10/f9rg6f.
(9) Mostofi, A. A.; Yates, J. R.; Pizzi, G.; Lee, Y.-S.; Souza, I.; Vanderbilt, D.; Marzari, N. An Updated Version of Wannier90: A Tool for Obtaining Maximally-Localised Wannier Functions. *Comput. Phys. Commun.* **2014**, *185* (8), 2309–2310. https://doi.org/10.1016/j.cpc.2014.05.003.
(10) Pizzi, G.; Vitale, V.; Arita, R.; Blügel, S.; Freimuth, F.; Géranton, G.; Gibertini, M.; Gresch, D.; Johnson, C.; Koretsune, T.; Ibañez-Azpiroz, J.; Lee, H.; Lihm, J.-M.; Marchand, D.; Marrazzo, A.; Mokrousov, Y.; Mustafa, J. I.; Nohara, Y.; Nomura, Y.; Paulatto, L.; Poncé, S.; Ponweiser, T.; Qiao, J.; Thöle, F.; Tsirkin, S. S.; Wierzbowska, M.; Marzari, N.; Vanderbilt, D.; Souza, I.; Mostofi, A. A.; Yates, J. R. Wannier90 as a Community Code: New Features and Applications. *J. Phys.: Condens. Matter* **2020**, *32* (16), 165902. https://doi.org/10/gg62rz.